\documentclass[twocolumn,english,pra,showpacs]{revtex4}
\usepackage[T1]{fontenc}
\usepackage[latin9]{inputenc}
\usepackage{color}
\usepackage{amsmath}
\usepackage{amssymb}
\usepackage{graphicx}
\usepackage{esint}

\makeatletter

\newcommand{\lyxdot}{.}

\@ifundefined{textcolor}{}
{%
 \definecolor{BLACK}{gray}{0}
 \definecolor{WHITE}{gray}{1}
 \definecolor{RED}{rgb}{1,0,0}
 \definecolor{GREEN}{rgb}{0,1,0}
 \definecolor{BLUE}{rgb}{0,0,1}
 \definecolor{CYAN}{cmyk}{1,0,0,0}
 \definecolor{MAGENTA}{cmyk}{0,1,0,0}
 \definecolor{YELLOW}{cmyk}{0,0,1,0}
}

\usepackage{babel}

\usepackage{babel}

\makeatother

\begin{document}

\title{Advantages of nonclassical pointer states in postselected weak measurements}

\author{Yusuf Turek$^{1,2}$}

\email{yusufu@itp.ac.cn}

\selectlanguage{english}%

\author{W.Maimaiti$^{2}$,Yutaka Shikano$^{3,4,5}$, Chang-Pu Sun$^{6}$}

\email{cpsun@csrc.ac.cn}

\selectlanguage{english}%

\author{M. Al-Amri$^{1,7}$}

\email{mdalamri@kacst.edu.sa}

\selectlanguage{english}%

\affiliation{$^{1}$The National Center for Applied Physics (NCAP), King Abdulaziz
City for Science and Technology (KACST), Riyadh 11442, Saudi Arabia}

\affiliation{$^{2}$State Key Laboratory of Theoretical Physics, Institute of
Theoretical Physics, Chinese Academy of Sciences, and University of
the Chinese Academy of Sciences, Beijing 100190, China}

\affiliation{$^{3}$Research Center of Integrative Molecular Systems (CIMoS),
Institute for Molecular Science, National Institutes of Natural Sciences,
Okazaki, Aichi 444-8585, Japan}

\affiliation{$^{4}$Institute for Quantum Studies, Chapman University, Orange,
CA 92866, USA.}

\affiliation{$^{5}$Materials and Structure Laboratory, Tokyo Institute of Technology,
Yokohama 226-8503, Japan.}

\affiliation{$^{6}$Beijing Computational Science Research Center, Beijing 100084,
China}

\affiliation{$^{7}$Institute for Quantum Studies and Department of Physics and
Astronomy, Texas A\&M University, College Station, Texas 77843-4242,
USA}
\begin{abstract}
We investigate, within the weak measurement theory, the advantages
of non-classical pointer states over semi-classical ones for coherent,
squeezed vacuum, and Schr\"{o}inger cat states. These states are utilized
as pointer state for the system operator $\hat{A}$ with property
$\hat{A}^{2}=\hat{I}$, where $\hat{I}$ represents the identity operator.
We calculate the ratio between the signal-to-noise ratio (SNR) of
non-postselected and postselected weak measurements. The latter is
used to find the quantum Fisher information for the above pointer
states. The average shifts for those pointer states with arbitrary
interaction strength are investigated in detail. One key result is
that we find the postselected weak measurement scheme for non-classical
pointer states to be superior to semi-classical ones. This can improve
the precision of measurement process. 
\end{abstract}


\maketitle

\section{Introduction}

The weak measurement, as a generalized von Neumann quantum measurement
theory, was proposed by Aharonov, Albert, and Vaidman\cite{Aharonov(1988)}.
In weak measurement, the coupling between pointer and measured systems
is sufficiently weak, but its induced weak value of the observable
on the measured system can be beyond the usual range of the eigenvalues
of that observable\cite{qparadox(2005)}. This feature of weak value
is usually referred to as an amplification effect for weak signals
rather than a conventional quantum measurement that collapses a coherent
superposition of quantum states\cite{Aharonov(1988),Tollaksen(2010)}.

After first optical implementation of weak value\cite{Ritchie(1991)},
it has been applied in different fields to observe very tiny effects,
such as beam deflection \cite{Pfeifer(2011),Hosten2008,Hogan(2011),Zhou(2013),Starling(2009),Dixon(2009)},
frequency shifts \cite{Starling(2010)-1}, phase shifts \cite{Starling(2010)},
angular shifts \cite{Magana(2013),Bertulio (2014)}, velocity shifts
\cite{viza(2013)}, and even temperature shift \cite{Egan(2012)}.
Weak value has a nature of being a complex number, which lead the
weak measurements to provide an ideal method to examine some fundamentals
of quantum physics. Quantum paradoxes (Hardy's paradox \cite{Aharonov(2002),Lundeen(2009),Yokota (2009)}
and the three-box paradox \cite{Resch(2004)}), quantum correlation
and quantum dynamics \cite{Aharonov(2005),Aharonov(2008),Holger(2013),Aharonov(2011),Shikano(2011),Shikano(2010)},
quantum state tomography \cite{Lundeen (2011),Lundeen(2012),Braveman(2013),Kocsis(2011),Malik(2014),Salvail(2013)},
violation of the generalized Leggett-Garg inequalities \cite{Palacios,Suzuki(2012),Dressel (2011),Emary(2014),Goggin(2011),Groen(2013)}
and violation of the initial Heisenberg measurement-disturbance relationship
\cite{Lee(2012),Eda(2014)} are just few examples. In these typical
examples, the small effects have been amplified due to the benefit
of weak values. This amplifying effect occurs when the preselection
and postselection states of the measured system are almost orthogonal.
The successful postselection probability tends to decrease in order
to have successful amplification effect. For more details about weak
measurement and weak value, one can consult these reviews \cite{Dressel(2014),Nori,Shikano(2012)}.

So far, most of weak measurement studies focus on using the zero-mean
Gaussian state as an initial pointer state. However, recent studies
\cite{Wu,Knee(2014)} have shown that zero-mean Gaussian pointer state
cannot improve the SNR when considering postselection probability.
Needless to say Gaussian beam is classical and one may naturally ask
how about using non-classical pointer states, and what kind of advantages
they have? This issue has been recently addressed \cite{Pang(2014)},
where coherent and coherent squeezed states were utilized as pointers.
They showed that the postselected weak measurement improved the SNR
compared to the non-postselected process if the pointer state, is
non-classical rather than classical. The focus of the calculation
was based on the assumption that the coupling between measuring device
and measured system is too weak, and hence it was enough to consider
the time evolution operator up to its first order. Furthermore, there
have been recent studies giving full order effects of the unitary
evolution due to the von Neumann interaction, but for classical and
semi-classical states\cite{Turek,Nakamura(2012)}.

In this paper, we address a remaining point of interest constructing
a general formula for weak measurement beyond the first order, and
utilizing the non-classical states. We investigate the advantages
of non-classical pointer states over classical (semi-classical ) pointer
state, within weak values, by considering postselection probability.
In order to do so, we use coherent, squeezed vacuum, and Schr\"{o}inger
cat states as pointer states for system observable $\hat{A}$ with
property $\hat{A}^{2}=\hat{I}$. We start by presenting an analytical
general expressions of the shifted values of position and momentum
operators for the above mentioned pointer states with arbitrary measurement
strengths. In addition, we present the ratio of SNR between postselected
and non-postselected weak measurement, and also look at quantum Fisher
information. Our key results in this paper are (i) Our general expressions
of shifted values reduce to the Nakamura's\cite{Nakamura(2012)} main
result if we take the zero-mean Gaussian beam as initial pointer state.
(ii) As shown in Ref.\cite{Pang(2014)}, improving the SNR using postselected
weak measurement, one needs the non-classical pointer states which
is better than classical or semi-classical states. (iii) Non-classical
pointer states are much better even when it comes to parameter estimation
process which is characterized by Fisher information.

The rest of the paper is organized as follows. In Section II, we give
the setup for our system. In Section III, we start by giving general
expressions for the expectation values of position and momentum operators.
After that we discuss the ratio of SNR between postselected and non-postselected
weak measurements of coherent, squeezed vacuum, and Schr\"{o}inger cat
states. In Section IV, we give the Fisher information for those given
states in the light of postselection probability. We give conclusion
to our paper in section V. Throughout this paper, we use the unit
$\hbar=1$.

\section{Setup}

For the weak measurement, the coupling interaction between system
and measuring device is given by the standard von Neumann Hamiltonian\cite{qparadox(2005)}

\begin{equation}
H=g\delta(t-t_{0})\hat{A}\otimes\hat{P}.\label{eq:Hamil}
\end{equation}
Here, $g$ is a coupling constant and $\hat{P}=\int p\vert p\rangle\langle p\vert dp$
is the conjugate momentum operator, while the position operator is
$\hat{X}=\int x\vert x\rangle\langle x\vert dx$ where $[\hat{X},\hat{P}]=i\hat{I}$.
We have taken, for simplicity, the interaction to be impulsive at
time $t=t_{0}$. For this kind of impulsive interaction the time evolution
operator becomes as $e^{-ig\hat{A}\otimes\hat{P}}$.

The weak measurement is characterized by the preselection and postselection
of the system state. If we prepare the initial state $\vert\psi_{i}\rangle$
of the system and the pointer state, and after some interaction time
$t_{0}$, we then postselect a system state $\vert\psi_{f}\rangle$
and obtain the information about a physical quantity $\hat{A}$ from
the pointer wave function by the following weak value: 
\begin{equation}
\langle A\rangle{}_{w}=\frac{\langle\psi_{f}\vert\hat{A}\vert\psi_{i}\rangle}{\langle\psi_{f}\vert\psi_{i}\rangle},\label{eq:WV}
\end{equation}
where the subscript $w$ denotes the weak value. From Eq. (\ref{eq:WV}),
we know that when the preselected state $\vert\psi_{i}\rangle$ and
the postselected state $\vert\psi_{f}\rangle$ are almost orthogonal,
the absolute value of the weak value can be arbitrarily large. This
feature leads to weak value amplification.

We express position operator $\hat{X}$ and momentum operator $\hat{P}$
in terms of the annihilation (creation) operator, $\hat{a}$($\hat{a}^{\dagger}$)
in Fock space representation as 
\begin{eqnarray}
\hat{X} & = & \sigma(\hat{a}^{\dagger}+\hat{a}),\label{eq:annix}\\
\hat{P} & = & \frac{i}{2\sigma}(\hat{a}^{\dagger}-\hat{a}),\label{eq:anniy}
\end{eqnarray}
where $\sigma$ is the width of the fundamental Gaussian beam. These
annihilation (creation) operators obey the commutation relation $[\hat{a},\hat{a}^{\dagger}]=\hat{I}$.
By substituting Eq.(\ref{eq:anniy}) into unitary evolution operator
$e^{-ig\hat{A}\otimes\hat{P}}$, bearing in mind that operator $\hat{A}$
satisfies the property $\hat{A}^{2}=\hat{I}$, we get: 
\begin{equation}
e^{-ig\hat{A}\otimes\hat{P}}=\frac{1}{2}(\hat{I}+\hat{A})\otimes D(\frac{s}{2})+\frac{1}{2}(\hat{I}-\hat{A})\otimes D(-\frac{s}{2}),\label{eq:UNA1}
\end{equation}
where parameter $s$ is defined by $s:\equiv g/\sigma$, and $D\left(\mu\right)$
is a displacement operator with complex number $\mu$ defined by 
\begin{equation}
D(\mu)=e^{\mu\hat{a}^{\dagger}-\mu^{\ast}\hat{a}}.\label{eq:DOP}
\end{equation}
Note that $s$ characterizes the measurement strength. Thus, we can
say that the coupling between system and pointer is weak (strong)\textcolor{black}{{}
and so the measurement is called weak (strong) measurement,} if $s\ll1$$(s\gg1)$.

\section{The shifted values and the signal-to-noise ratio (SNR) }

In this section we start by giving general shifted values of semi-classical
state (coherent state) and non-classical states; squeezed vacuum and
Schr\"{o}inger cat pointer states for arbitrary measurement strength
$s$. To show the advantages of non-classical pointer states over
semi-classical ones, we discuss the ratio of SNR between postselected
and non-postselected weak measurements

\begin{equation}
\mathcal{\chi=}\frac{R_{X}^{p}}{R_{X}^{n}}.\label{eq:Ratio}
\end{equation}
Here, $R_{X}^{p}$ represents the SNR of postselected weak measurement
defined as 
\begin{equation}
R_{X}^{p}=\frac{\sqrt{NP_{s}}\vert\langle X\rangle_{fi}\vert}{\sqrt{\langle X^{2}\rangle_{f}-\langle X\rangle_{f}^{2}}}.\label{eq:SNR}
\end{equation}
Here, $N$ is the total number of measurements, $P_{s}$ is probability
of finding the postselected state for a given preselected state, and
$NP_{s}$ is the number of times the system was found in a postselected
state. Here, $\langle\rangle_{f}$ denotes the expectation value of
measuring observable under the final state of the pointer.

When dealing with non-postselected measurement, there is no postselection
process after the interaction between system and measuring device
due to unitary evolution operator $e^{-ig\hat{A}\otimes\hat{P}}$.
Therefore, the definition of $R_{X}^{p}$ for non-postselected weak
measurement can be given as 
\begin{equation}
R_{X}^{n}=\frac{\sqrt{N}\vert\langle X\rangle_{f^{\prime}i}\vert}{\sqrt{\langle X^{2}\rangle_{f^{\prime}}-\langle X\rangle_{f^{\prime}}^{2}}}.
\end{equation}
Here, $\langle\rangle_{f^{\prime}}$ denotes the expectation value
of measuring observable under the final state of the pointer without
postselection.

\subsection{Coherent pointer state }

Coherent state is typical semi-classical state which satisfies the
minimum Heisenberg uncertainty relation. Here, we take the coherent
state \cite{Scully} as initial pointer state 
\begin{equation}
\left|\alpha\right\rangle =D(\alpha)\left|0\right\rangle ,\label{eq:Coh sta}
\end{equation}
where $\alpha=re^{i\phi}$ is an arbitrary complex number. After unitary
evolution given in Eq. (\ref{eq:UNA1}), the resultant system state
is postselected to $\vert\psi_{f}\rangle$. Then, we obtain the following
normalized final pointer state:

\begin{eqnarray}
\vert\Psi_{f_{1}}\rangle & \!\!\!\!= & \frac{\lambda}{2}\times\label{eq:GeC}\\
 & \!\! & \!\!\!\!\!\![(1+\!\!\langle A\rangle_{w})e^{-i\frac{s}{2}\Im\left(\alpha\right)}\vert\alpha\!\!+\frac{s}{2}\rangle+(1-\!\!\langle A\rangle_{w})e^{i\frac{s}{2}\Im\left(\alpha\right)}\vert\alpha\!\!-\frac{s}{2}\rangle],\nonumber 
\end{eqnarray}
where the normalization coefficient is given as 
\begin{eqnarray}
\lambda & = & \sqrt{2}\times\label{eq:Co1}\\
\!\! & \!\!\!\!\!\! & [1\!+\!\vert\langle A\rangle_{w}\vert^{2}\!\!+\!\!\Re((1\!-\!\langle A\rangle{}_{w}^{\ast})(1\!\!+\!\!\langle A\rangle_{w})e^{-2is\Im\left(\alpha\right)})e^{-\frac{1}{2}s^{2}}]^{-\frac{1}{2}},\nonumber 
\end{eqnarray}
and $\Im$ ($\Re$) represents the imaginary (real) part of a complex
number. Using Eqs.(\ref{eq:GeC},\ref{eq:Co1}) we can calculate general
forms of the expectation values of conjugate position operator $X$
and momentum operator $P$, under the final pointer state $\vert\Psi_{f_{1}}\rangle$,
to be 
\begin{eqnarray}
\langle X\rangle_{f_{1}} & = & \sigma\vert\lambda\vert^{2}\{(1+\vert A{}_{w}\vert^{2})\Re(\alpha)+s\Re\langle A\rangle_{w}\label{eq:X1}\\
 &  & \!\!\!\!+\Re[(1\!\!-\!\!\langle A\rangle_{w}^{\ast})(1\!\!+\!\!\langle A\rangle_{w})e^{-2si\Im(\alpha)}]\Re(\alpha)e^{-\frac{1}{2}s^{2}}\}\nonumber 
\end{eqnarray}
and

\begin{eqnarray}
\langle P\rangle_{f_{1}} & = & \frac{\vert\lambda\vert^{2}}{4\sigma}\{2(1+\vert\langle A\rangle_{w}\vert^{2})\Im(\alpha)\label{eq:P1}\\
 &  & \!\!\!\!-\Im[(1-\!\!\langle A\rangle_{w})(1+\!\!\langle A\rangle_{w}^{\ast})e^{2is\Im(\alpha)}(s-2i\Im(\alpha))]e^{-\frac{1}{2}s^{2}}\!\!\},\nonumber 
\end{eqnarray}
respectively. Eqs. (\ref{eq:X1}, \ref{eq:P1}) are the general forms
of expectation values for system operator $\hat{A}$, with the property
$\hat{A}^{2}=\hat{I}$, and they are valid for any arbitrary value
of the measurement strength parameter $s$.

Here, we assume that the operator to be observed is the spin $x$
component of a spin- $1/2$ particle through the von Neuman interaction
\begin{equation}
A=\sigma_{x}=\vert\uparrow_{z}\rangle\langle\downarrow_{z}\vert+\vert\downarrow_{z}\rangle\langle\uparrow_{z}\vert,
\end{equation}
where $\vert\uparrow_{z}\rangle$ and $\langle\downarrow_{z}\vert$
are eigenstates of $\sigma_{z}$ with corresponding eigenvalues $1$
and $-1$, respectively. When we select the preselected and postselected
states as 
\begin{equation}
\vert\psi_{i}\rangle=\cos(\frac{\theta}{2})\vert\uparrow_{z}\rangle+e^{i\varphi}\sin(\frac{\theta}{2})\vert\downarrow_{z}\rangle,\label{eq:Pre}
\end{equation}
and 
\begin{equation}
\vert\psi_{f}\rangle=\vert\uparrow_{z}\rangle,\label{eq:Post}
\end{equation}
respectively, we can get the weak value by substituting these states
to 
\begin{equation}
\langle A\rangle_{w}=\langle\sigma_{x}\rangle_{w}=\frac{\langle\psi_{f}\vert A\vert\psi_{i}\rangle}{\langle\psi_{f}\vert\psi_{i}\rangle},
\end{equation}
obtaining 
\begin{equation}
\langle A\rangle_{w}=e^{i\varphi}\tan(\frac{\theta}{2}).\label{eq:WV-1}
\end{equation}
where, $\theta\in[0,\pi]$ and $\varphi\in[0,2\pi)$. Here, the postselection
probability is $P_{s}=\cos^{2}(\frac{\theta}{2})$.\textcolor{black}{{}
Throughout this paper, we use the above preselected and postselected
states and weak value, which are given in Eq.(\ref{eq:Pre},\ref{eq:Post})
and Eq.(\ref{eq:WV-1}) for our discussions.}\textcolor{red}{{} }

\begin{figure}
\includegraphics{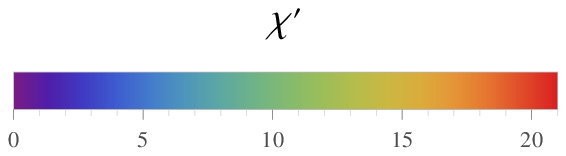}

\includegraphics[width=8cm]{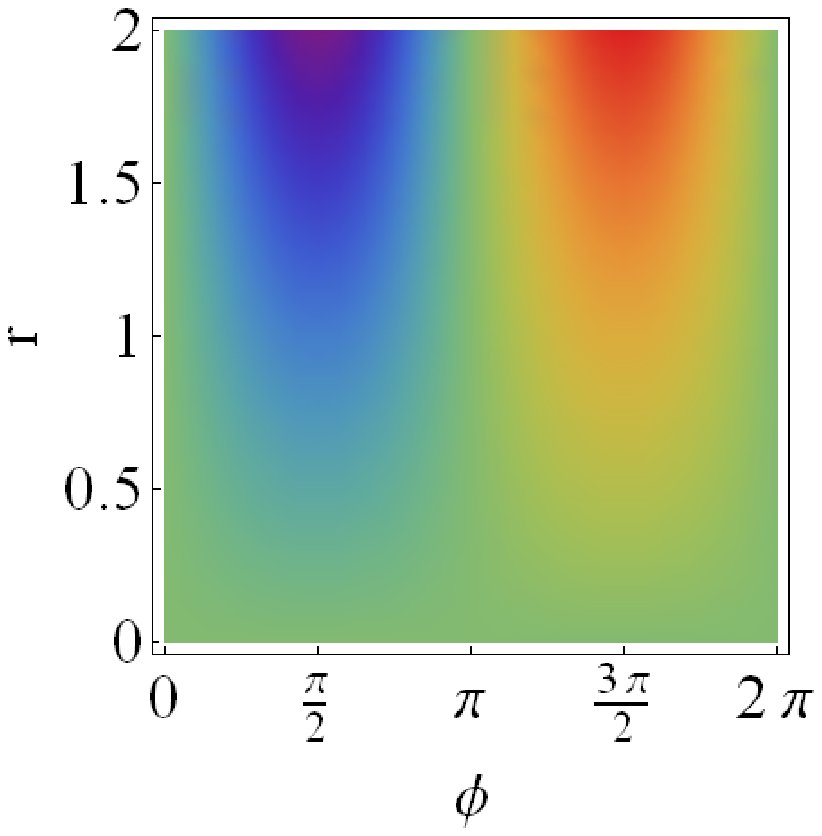}

\protect\caption{\label{fig:1}(Color online)The ratio $\chi^{\prime}$ of SNRs between
postselected and non-postselected weak measurement vs coherent state's
parameters $\phi$ and r . Here we take $\varphi=\pi/4$ , $\theta=7\pi/9$
, and $s=10^{-5}$.}
\end{figure}

In the case of coherent state is used as initial pointer state, we
calculate the SNR of postselected and non-postselected process in
weak measurement regime ($s\ll1$). In Fig. \ref{fig:1} we plot the
ratio $\chi^{\prime}=(\chi-1.4618)\times10^{5}$ against coherent
state's parameters $r$ and $\phi$, where the ratio $\chi$ has the
same value1.4618 in most of the regions. This means that, for coherent
state pointer, the postselected weak measurement is little better
than non-postselected case which in turn slightly increase the precision
of measurement.

\subsection{Squeezed vacuum state }

Squeezed vacuum state is a typical quantum state. It has many applications
in optical communication, optical measurement, and gravitational wave
detection \cite{Milburn}. Here, we assume that the initial pointer
is squeezed vacuum state \cite{Scully} which is defined by 
\begin{equation}
\vert\xi\rangle=S(\xi)\left|0\right\rangle .\label{eq:Ss}
\end{equation}
Here, 
\begin{equation}
S(\xi)=\exp(\frac{1}{2}\xi^{\ast}a^{2}-\frac{1}{2}\xi a^{\dagger2}),\label{eq:ss1}
\end{equation}
where the squeezing parameter $\xi=\eta e^{i\delta}$ is an arbitrary
complex number. After unitary evolution given in Eq. (\ref{eq:UNA1}),
the total system state is postselected to $\vert\psi_{f}\rangle$.
Then, we obtain the following normalized final pointer state: 
\begin{equation}
\vert\Psi_{f_{2}}\rangle=\frac{\gamma^{\prime}}{2}[(1+\langle A\rangle_{w})\vert\frac{s}{2},\xi\rangle+(1-\langle A\rangle_{w})\vert-\frac{s}{2},\xi\rangle],
\end{equation}
where the normalization coefficient is given by 
\begin{equation}
\gamma^{\prime}=\sqrt{2}[1+\vert\langle A\rangle_{w}\vert^{2}+(1-\vert\langle A\rangle_{w}\vert^{2})e^{-\frac{1}{2}s^{2}\vert\cosh\eta+e^{i\delta}\sinh\eta\vert^{2}}]^{-\frac{1}{2}}
\end{equation}
and we note that $\vert\pm\frac{s}{2},\xi\rangle=D(\pm\frac{s}{2})S(\xi)\vert0\rangle$
is squeezed coherent state. Next we will calculate the expectation
values of position and momentum operators under the normalized final
pointer state $\vert\Psi_{f_{2}}\rangle$, and the results reads 
\begin{eqnarray}
\langle X\rangle_{f_{2}} & \!\!= & g\vert\gamma^{\prime}\vert^{2}\Re\langle A\rangle_{w}\\
 &  & -g\vert\gamma^{\prime}\vert^{2}\Im\langle A\rangle_{w}e^{-\frac{1}{2}s^{2}\vert\cosh\eta+e^{i\delta}\sinh\eta\vert^{2}}\sinh(2\eta)\sin\delta\nonumber 
\end{eqnarray}
and 
\begin{eqnarray}
\langle P\rangle_{f_{2}} & = & \frac{g\vert\gamma^{\prime}\vert^{2}}{2\sigma^{2}}\times\\
 &  & \Im\langle A\rangle_{w}e^{-\frac{1}{2}s^{2}\vert\cosh\eta+e^{i\delta}\sinh\eta\vert^{2}}(1+\sinh\left(2\eta\right)\cos\delta),\nonumber 
\end{eqnarray}
respectively. These formulas are valid not only in the weak measurement
regime ($s\ll1$), but also in strong measurement regime ($s\gg1$).

\begin{figure}
\includegraphics{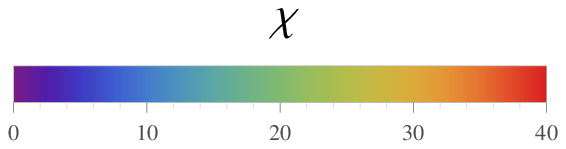}

\includegraphics[width=8cm]{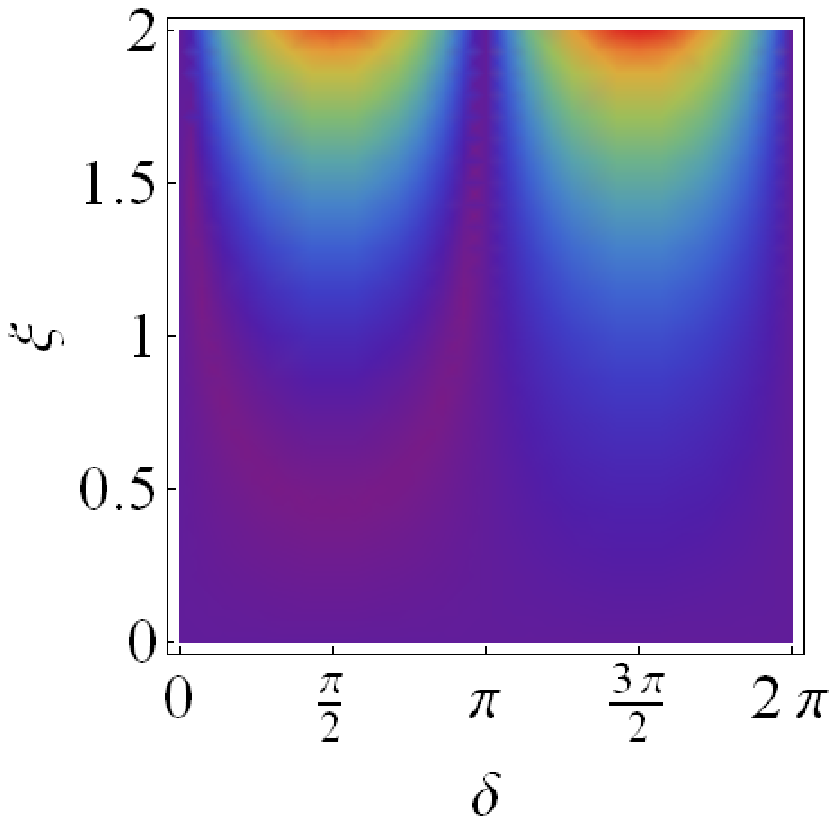}

\protect\protect\caption{(Color online)\label{fig:2} The ratio $\mathcal{\chi}$ of SNRs between
postselected and non-postselected weak measurements vs squeezed vacuum
state's parameters $\delta$ and $\eta$. Here we take $\varphi=\pi/4$,
$\theta=7\pi/9$, and $s=10^{-5}$.}
\end{figure}

Fig.\ref{fig:2} shows the ratio $\mathcal{\chi}$ of SNR for squeezed
pointer state between postselected and non- postselected weak measurements
($s\ll1$) plotted as a function of $\delta$ and $\eta$ which are
the parameters of squeezed state. One can see that when $\eta$ is
large and near the points where $\delta=\frac{\pi}{2},\frac{3\pi}{2}$
the ratio $\mathcal{\chi}$ is much larger than unity. Evidently,
this result indicates that squeezed pointer state is one of the quantum
state candidates that can be utilized to improve the SNR in postselected
rather than non-postselected weak measurement. This result was also
confirmed in Ref. \cite{Pang(2014)}.

\subsection{Schr\"{o}inger cat state }

Schr\"{o}inger cat state is another typical quantum state\cite{Neito}
which is a superposition of two coherent correlated states moving
in opposite directions. Generally, there are two kinds of Schr\"{o}inger
cat states\cite{Dodonov}; even and odd Schr\"{o}inger cat states. Even
Schr\"{o}inger cat state has very similar properties with squeezed state
\cite{Hollen}, since it has superpositions of photon number states
with even numbers of quanta. Therefore, we consider the even Schr\"{o}inger
cat state as initial pointer sate to examine further the advantages
of non-classical pointer state. The normalized even Schr\"{o}inger cat
state can be written as

\begin{equation}
\vert\Theta_{+}\rangle=K(\vert\alpha\rangle+\vert-\alpha\rangle),
\end{equation}
where $\vert\pm\alpha\rangle$ are coherent states as defined in Eq.
(\ref{eq:Coh sta}) which is characterized by $\alpha=re^{i\phi}$,
and the normalization constant is 
\begin{equation}
K=\frac{1}{\sqrt{2+2e^{-2\vert\alpha\vert^{2}}}}.
\end{equation}
Following the same procedure as in previous sections, after taking
unitary evolution given in Eq. (\ref{eq:UNA1}), the outcome will
then be projected to postselected state, $\vert\psi_{f}\rangle$.
Then, we obtain the following normalized final pointer state 
\begin{equation}
\vert\Psi_{f_{3}}\rangle=\frac{\kappa^{\prime}}{2}[(1+\langle A\rangle_{w})D(\frac{s}{2})+(1-\langle A\rangle_{w})D(-\frac{s}{2})]\vert\Theta_{+}\rangle,\label{eq:SS}
\end{equation}
where the normalization coefficient is given by 
\begin{eqnarray}
\kappa^{\prime} & = & [\frac{1}{2}(1+\vert\langle A\rangle_{w}\vert^{2})+K^{2}(1-\vert\langle A\rangle_{w}\vert^{2})\cos\left(2s\Im\left(\alpha\right)\right)e^{-\frac{s^{2}}{2}}\nonumber \\
 &  & +\frac{K^{2}}{2}(1-\vert\langle A\rangle_{w}\vert^{2})(e^{-\frac{1}{2}\vert2\alpha+s\vert^{2}}+e^{-\frac{1}{2}\vert2\alpha-s\vert^{2}})]^{-\frac{1}{2}}.
\end{eqnarray}
By using Eq.(\ref{eq:SS}) we calculate, in straightforward manner,
the general forms of the expectation values for both conjugate position
and momentum operators as 
\begin{eqnarray}
\langle X\rangle_{f_{3}} & = & 2\sigma\vert\kappa^{\prime}\vert^{2}K^{2}\times\{s\Re\langle A\rangle_{w}(1\!\!+e^{-2\vert\alpha\vert^{2}})\!\\
 &  & \!\!\!\!\!+2\Im\langle A\rangle_{w}\Re\left(\alpha\right)\sin(2s\Im\left(\alpha\right)e^{-\frac{1}{2}s^{2}}\nonumber \\
 &  & -\Im\langle A\rangle_{w}\Im\left(\alpha\right)(e^{-\frac{1}{2}\vert2\alpha+s\vert^{2}}-e^{-\frac{1}{2}\vert2\alpha-s\vert^{2}}))\}\nonumber 
\end{eqnarray}
and 
\begin{eqnarray}
\langle P\rangle_{f_{3}} & = & \frac{\vert\kappa\vert^{2}K^{2}\Im\langle A\rangle_{w}}{2\sigma}\times\{\left(2\Re\left(\alpha\right)+s\right)e^{-\frac{1}{2}\vert2\alpha+s\vert^{2}}\nonumber \\
 &  & \!\!\!\!+\!4\sin\left[2s\Im\left(\alpha\right)\right]\Im\left(\alpha\right)e^{-\frac{1}{2}s^{2}}\!\!\!\!+\!2s\cos\left[2s\Im\left(\alpha\right)\right]e^{-\frac{1}{2}s^{2}}\nonumber \\
 &  & \!\!\!\!-\left(2\Re\left(\alpha\right)-s\right)e^{-\frac{1}{2}\vert2\alpha-s\vert^{2}}\},
\end{eqnarray}
respectively.

\begin{figure}
\includegraphics{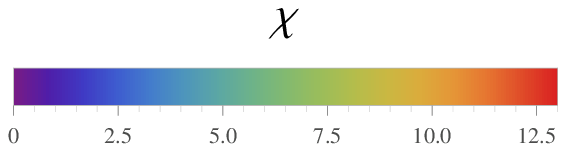}

\includegraphics[width=8cm]{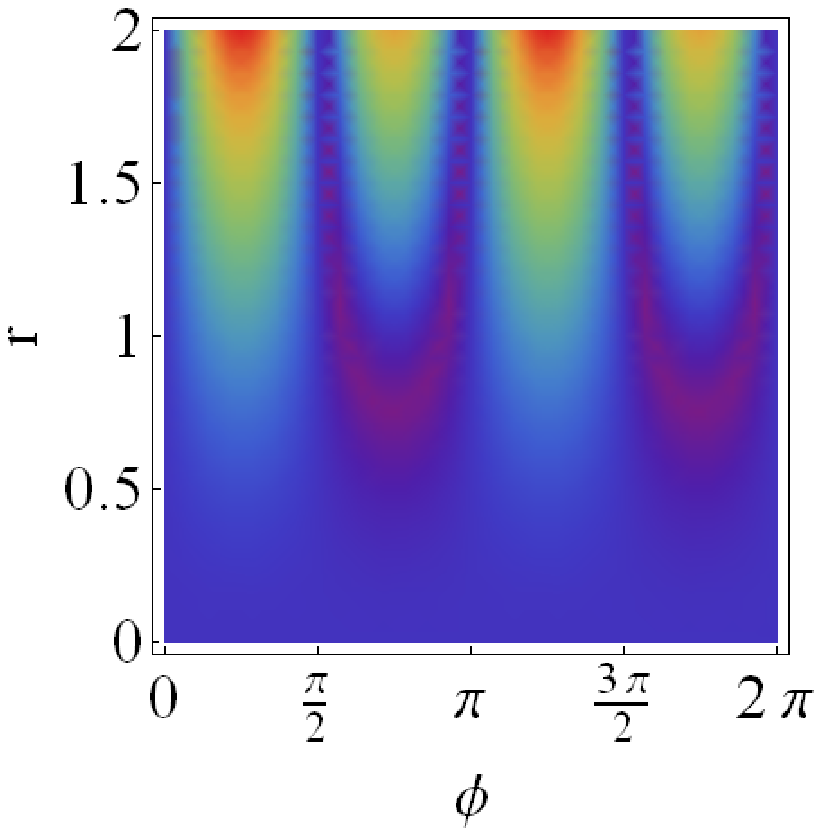}

\protect\protect\caption{(Color online) \label{fig:3}The ratio $\mathcal{\chi}$ of SNRs between
postselected and non post-selected measurement vs Schr\"{o}inger cat
state's parameters $\phi$ and $r$. Here we take $\varphi=\pi/4$,$\theta=7\pi/9$,
and $s=10^{-5}$.}
\end{figure}

In Fig. \ref{fig:3}, we plot the ratio $\chi$ of SNRs between postselected
and non-postselected weak measurements for Schr\"{o}inger cat pointer
state. It is, clearly, indicating that when $r$ is increased and
passed near $\phi=\frac{\pi}{4},\frac{3\pi}{4},\frac{5\pi}{4},\frac{7\pi}{4}$
, the ratio of SNRs is much larger than unity. Furthermore, when comparing
Fig. \ref{fig:3} to Fig. \ref{fig:1}, we find that the ratio $\chi$of
non-classical Schr\"{o}inger cat pointer state is higher than semi-classical
coherent pointer state for the same parameters. This, evidently, leads
to the improvement of SNR. However, when comparing between the two
non-classical states in Fig. \ref{fig:2} and Fig. \ref{fig:3}, one
can see that these two Figures have some similarity, where both of
them have the ratio $\chi$ larger than unity, while it get much stronger
value for the case of squeezed state.

We have to emphasize at this point that we have also calculated the
odd Schr\"{o}inger cat pointer states but found that they have similar
properties and results like the even Schr\"{o}inger cat pointer states.
And in order to avoid repetition, therfore, we just report the results
of the even Schr\"{o}inger cat states.

For the ratio of SNRs between postselected and non-postselected weak
measurements, we can conclude that non-classical pointer states (squeezed
vacuum, and Schr\"{o}inger cat state) are better than semi-classical
one (coherent sate) in order to improve the SNR in postselected weak
measurements ($s\ll1$) for complex weak values. This conclusion can
be seen clearly from Fig. \ref{fig:1}, Fig. \ref{fig:2} and Fig.
\ref{fig:3}.

The general expectation values of position and momentum operators
for the above three pointer states - coherent, squeezed, and Schr\"{o}inger
cat states - can have the same property. This can be achieved if we
assume the initial pointer state to be a zero-mean Gaussian beam (This
corresponding to $r=0$ for coherent states and Schr\"{o}inger cat state,
and $\eta=0$ for squeezed vacuum state, respectively ), then all
expressions reduced to 
\begin{equation}
\langle X\rangle_{f}=\frac{g\Re\langle A\rangle_{w}}{\mathcal{Z}},\label{eq:RedX}
\end{equation}
and 
\begin{equation}
\langle P\rangle_{f}=\frac{g\Im\langle A\rangle_{w}}{2\sigma^{2}\mathcal{Z}}e^{-\frac{1}{2}s^{2}}.\label{eq:RedP}
\end{equation}
Here, 
\begin{equation}
\mathcal{Z=}1+\frac{1}{2}(1-\vert\langle A\rangle_{w}\vert^{2})(e^{-\frac{1}{2}s^{2}}-1).
\end{equation}
This result is given in Nakamura's work \cite{Nakamura(2012)}.

\begin{widetext}

\begin{figure}
\includegraphics{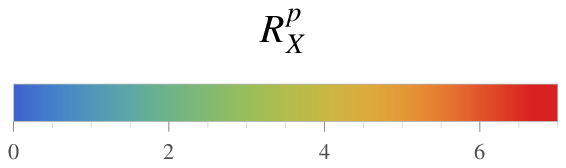}

\includegraphics[width=5.3cm]{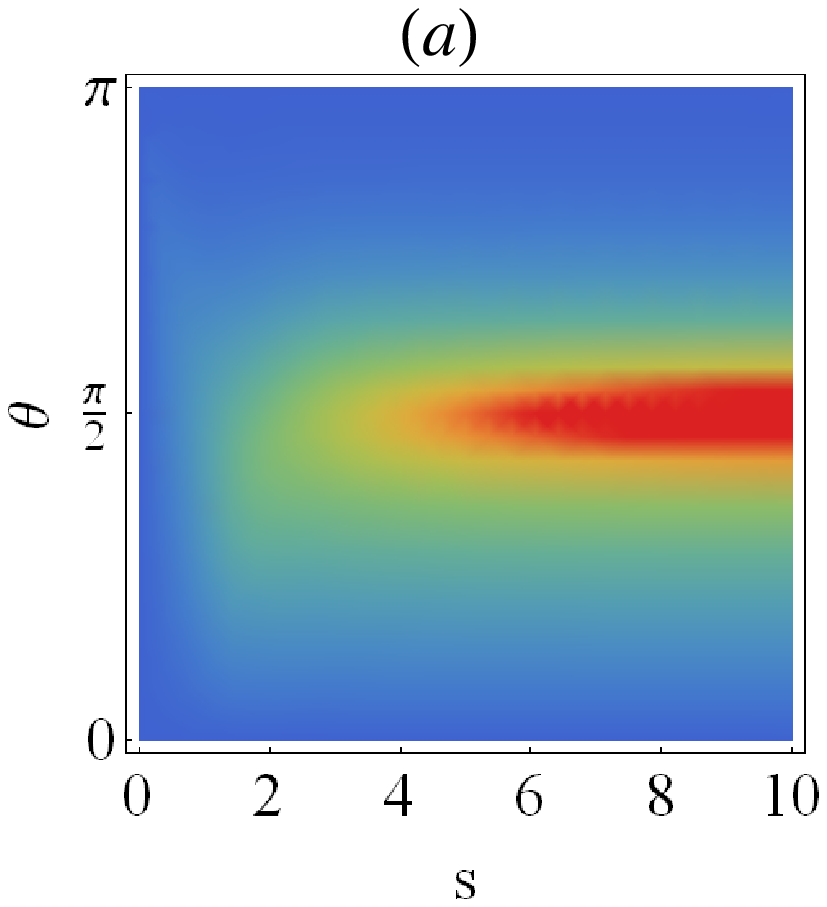}\includegraphics[width=5.3cm]{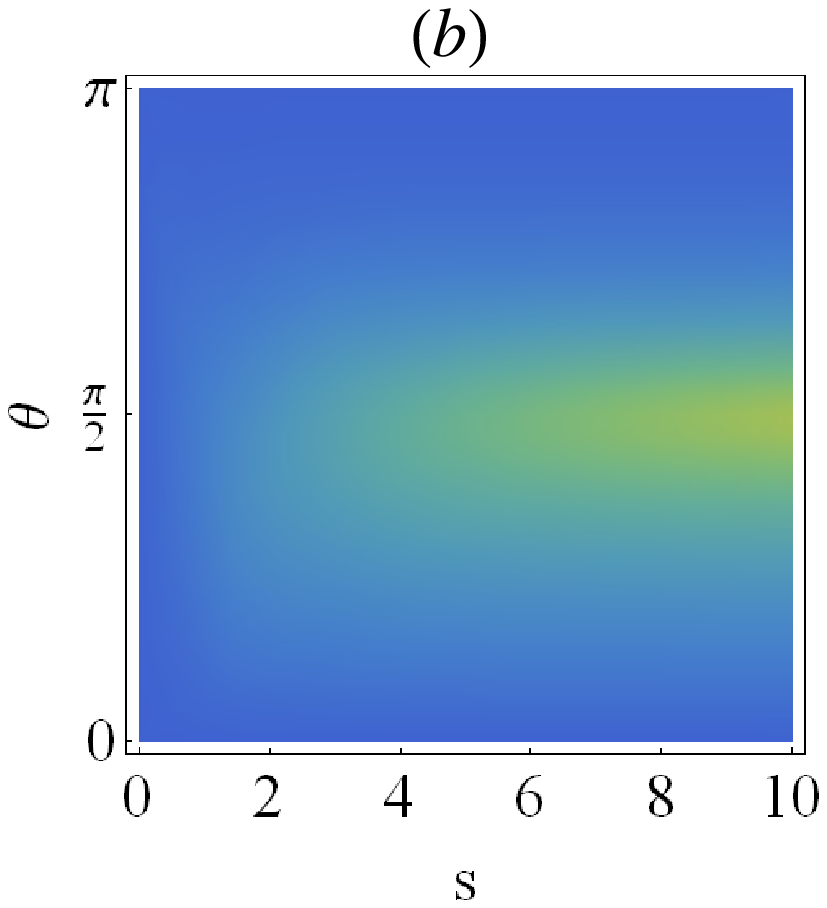}\includegraphics[width=5.3cm]{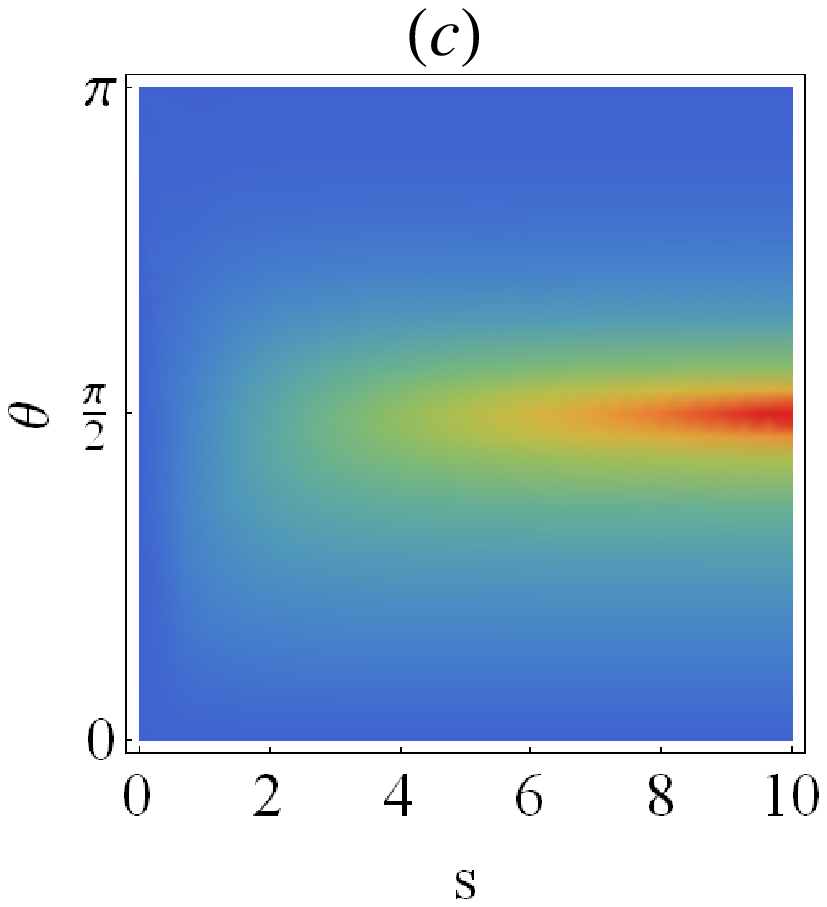}

\protect\protect\caption{(Color online) \label{fig:labc}The $R_{X}^{p}$ for arbitrary measurement
strength parameter $s$, and $\theta$ for different weak values.
Here, we take $\varphi=0$ and $N=1$. (a) For coherent pointer state,
$r=1$, $\phi=\pi/4$. (b) For Schr\"{o}inger cat pointer sates, $r=1,$
$\phi=\pi/4$. (c) For squeezed vacuum pointer state, $\eta=1,$$\delta=\pi/4$. }
\end{figure}

\end{widetext}

A remaining issue is to examine the connection between weak and strong
postselected measurement. Thus, we plot the $SNR_{X}^{p}$, which
is defined in Eq.(\ref{eq:SNR}), as function of arbitrary measurement
strength parameter $s$ and preselection angle $\theta$. From Fig.
\ref{fig:labc}, particularly for squeezed vacuum pointer state, we
can see that at $\theta=\pi/2$ the $SNR_{X}^{p}$ increase with the
increase of $s$, this is the strong measurement result. The reason
is that at $\theta=\pi/2$ the preselected state Eq.(\ref{eq:Pre})
is the eigenstate of operator $\sigma_{x}$ which have eigenvalue
$+1$. This figure doesn't only make the connection between weak and
strong postselected measurement, but also indicates that non-classical
pointer states are also good enough comparing with semi-classical
ones in generalized von Neumann measurement\cite{Turek}.

\section{Quantum Fisher information}

Fisher information is the maximum amount of information about the
parameter that we can extract from the system. For a pure quantum
state $\vert\psi_{s}\rangle$, the quantum Fisher information estimating
$s$ is 
\begin{equation}
F^{(Q)}=4[\langle\partial_{s}\psi_{s}\vert\partial_{s}\psi_{s}\rangle-\vert\langle\psi_{s}\vert\partial_{s}\psi_{s}\rangle\vert^{2}].
\end{equation}
where the state $\vert\psi_{s}\rangle$ represents the final pointer
states of the system. Here, this can be used for coherent, squeezed
vacuum, or Schr\"{o}inger cat states when only dealing with the postselected
weak measurement in the first order evolution of unitary operator
$e^{-ig\hat{A}\otimes\hat{P}}$. Here, $s\equiv g/\sigma$ is the
measurement strength parameter which directly related to coupling
constant $g$ in our Hamiltonian of Eq.(\ref{eq:Hamil}).

The variance of unknown parameter $\Delta s$ is bounded by the Cramer-Rao
bound 
\begin{equation}
\Delta s\geq\frac{1}{NF^{(Q)}},
\end{equation}
where $N$ is the total number of measurements. Thus, the Fisher information
set the minimal possible estimate for parameter $s$, while higher
Fisher information means better estimation. In weak measurement, if
we consider the successful postselection probability, then Fisher
information would be $F_{p}^{(Q)}=P_{s}F^{(Q)}$. In Ref.\cite{Pang(2014)},
one can find general proof showing that quantum Fisher information
is higher in postselected rather than non-postselected weak measurement.
Thus, we just focus on the postselected weak measurement process and
look into Fisher information for semi-classical and non-classical
pointer states.

\begin{widetext}

\begin{figure}
\includegraphics{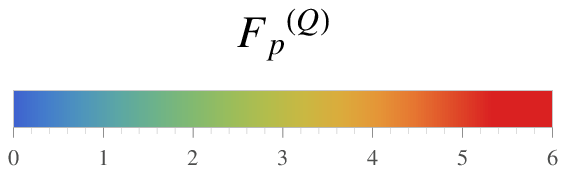}

\includegraphics[width=5.3cm]{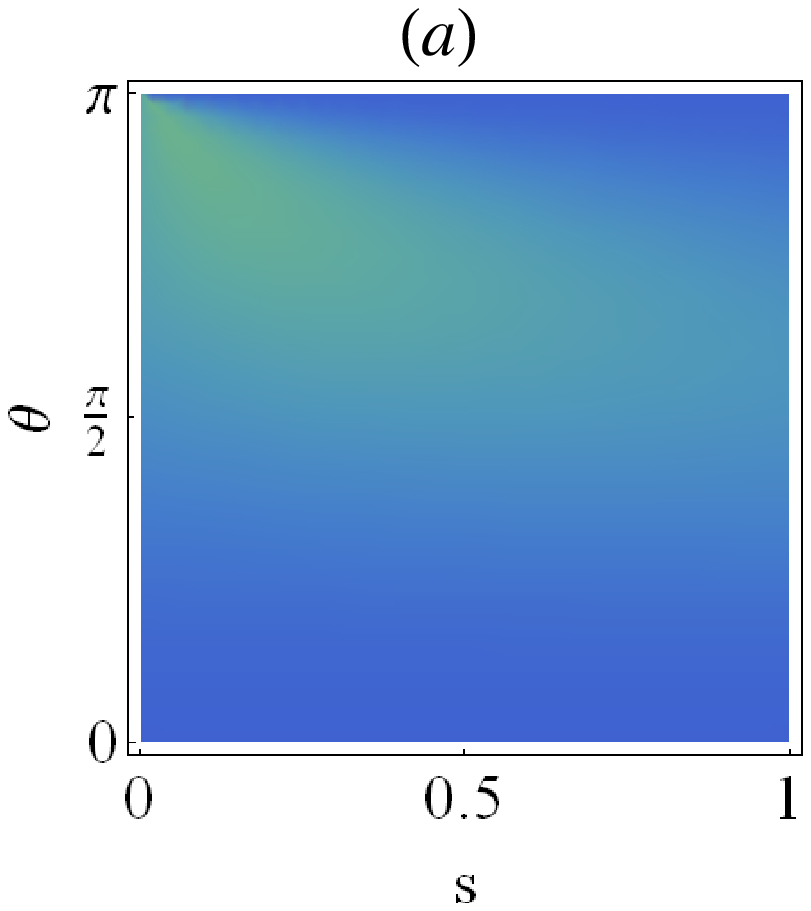}\includegraphics[width=5.3cm]{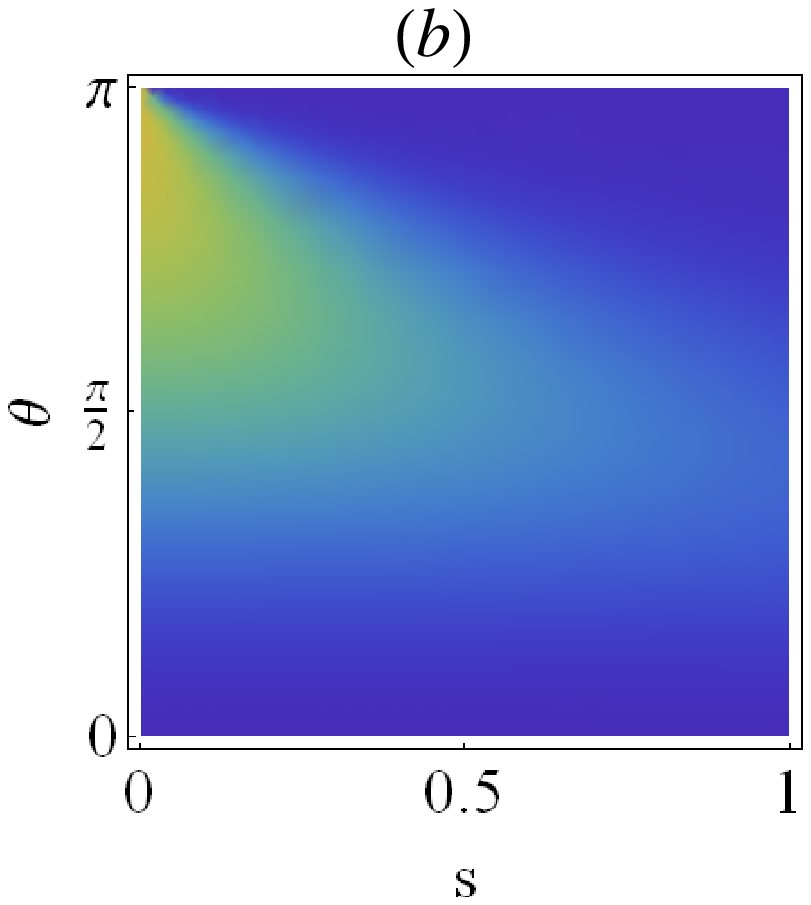}\includegraphics[width=5.3cm]{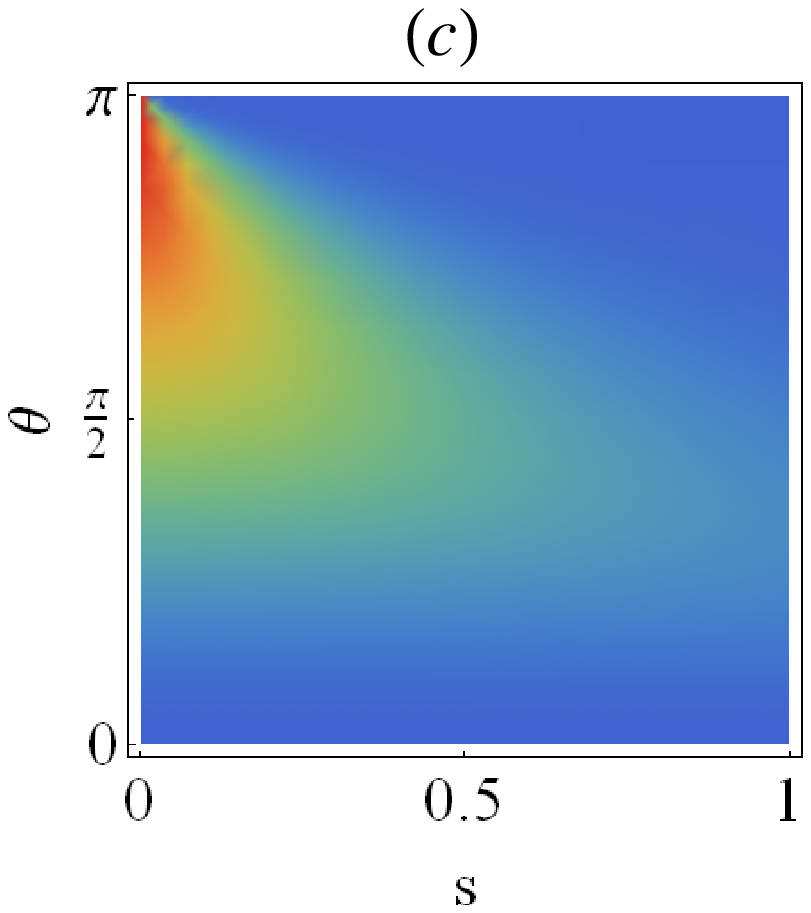}

\protect\protect\caption{(Color online) \label{fig:4}The quantum Fisher information in weak
measurement regime vs measurement strength parameter $s$ and $\theta$
for different weak values. Here, we take $\varphi=\pi/4$ and $N=1$.
(a) For coherent pointer state, $r=1$, $\phi=\pi/4$. (b) Schr\"{o}inger
cat pointer sates, $r=1,$ $\phi=\pi/4$. (c) For squeezed vacuum
pointer state, $\eta=1,$$\delta=\pi/4$. }
\end{figure}

\end{widetext}

We proceed investigating the variation of Fisher information in weak
measurement regime for different weak values. Our numerical results
in Fig. \ref{fig:4} show that the quantum Fisher information is higher
in weak measurement regime ($s\ll1$) when the preselection and postselection
state almost orthogonal. The other important result is that the non-classical
pointer states have more advantages over the semi-classical ones which
in turn leads to better estimation process.

\section{Conclusion }

In summary, we give a general expressions for the shifted values of
position and momentum operators for different pointer states (coherent,
squeezed vacuum, and Schr\"{o}inger cat states), these expressions are
valid in weak and strong measurement regimes. In the next step, we
investigate the SNR and the quantum Fisher information only in weak
measurement regime. We find that if we take initial state as zero
mean Gaussian state, our general expressions of shifted values would
be reduced to Eq. (\ref{eq:RedX}) and Eq. (\ref{eq:RedP}), which
are given in Ref. \cite{Nakamura(2012)}. By giving the ratio of SNR
between postselected and non- postselected weak measurement, we find
that postselected weak meaurement process for non-classical pointer
states give more infromation about the system comparing to the non-postselected
process. This result keeps consistent with S. Pang et al's work \cite{Pang(2014)}.
If one wants to quantify the quantum Fisher information in order to
improve the precision of unknown parameter estimation, then he can
consider using non-classical pointer sate and avoiding the semi-classical
one. 
\begin{acknowledgments}
Y.T would like to thank Yinan Fang for useful suggestions and discussions.
Y.S. thanks the financial supports by the Center for the Promotion
of Integrated Sciences (CPIS) of Sokendai, ICRR Joint Research from
The University of Tokyo, and NINS Youth Collaborative Project to cooperate
in this project.This research is supported by a grant from King Abdulaziz
City for Science and Technology (KACST).
\end{acknowledgments}

\end{document}